\documentclass[fdp,fleqn]{w-art}
\usepackage{times}
\usepackage{w-thm}
\usepackage{amsmath}
\usepackage{amscd}
\usepackage{amsfonts}
\usepackage[]{graphicx}
\newcommand{\be}{\begin{eqnarray}}
\newcommand{\ee}{\end{eqnarray}}
\newcommand{\nn}{\nonumber}

\newcommand{\cV}{{\cal V}}
\newcommand{\cQ}{{\cal Q}}
\newcommand{\cP}{{\cal P}}
\newcommand{\p}{\partial}

\newcommand{\eps}{\epsilon}

\newcommand{\beq}{\begin{equation}}
\newcommand{\eeq}{\end{equation}}

\newcommand{\f}{\frac}
\newcommand{\mc}{\mathcal}
\newcommand{\hs}{\hspace{0.1 cm}}

\newcommand{\mf}{\mathfrak}
\newcommand{\mbb}{\mathbb}

\newcommand{\om}{\omega}

\newcommand{\pa}{\partial}

\begin{document}
\DOIsuffix{theDOIsuffix}
\Volume{55}
\Issue{1}
\Month{01}
\Year{2007}
\pagespan{1}{}

\begin{flushright}
ULB-TH/09-04
\end{flushright}

\title[Massive IIA Supergravity and $E_{10}$]{Massive Type IIA Supergravity and $E_{10}$}


\author[M. Henneaux]{Marc Henneaux\inst{1,2,}
  \footnote{henneaux@ulb.ac.be}}
\address[\inst{1}]{Physique Th\'eorique et Math\'ematique,\\
Universit\'e Libre
de Bruxelles \& International Solvay Institutes,\\ 
ULB-Campus Plaine C.P. 231, B-1050 Bruxelles, Belgium.}
\address[\inst{2}]{Centro de Estudios Cient\'{\i}ficos
(CECS), Casilla 1469, Valdivia, Chile.}
\author[E. Jamsin]{Ella Jamsin\inst{1,}\footnote {ejamsin@ulb.ac.be}}
\author[A. Kleinschmidt]{Axel Kleinschmidt\inst{1,}\footnote{akleinschmidt@ulb.ac.be}}
\author[D. Persson]{Daniel Persson\inst{1,3,}\footnote{dpersson@ulb.ac.be}}
\address[\inst{3}]{Fundamental Physics, Chalmers University of Technology, SE-412 96, G\"oteborg, Sweden.}
\begin{abstract}

In this talk we investigate the symmetry under $E_{10}$ of Romans' massive type IIA supergravity. We show that the dynamics of a spinning particle in a non-linear sigma model on the coset space $E_{10}/K(E_{10})$ reproduces the bosonic and fermionic dynamics of massive IIA supergravity, in the standard truncation. In particular, we identify Romans' mass with a generator of $E_{10}$ that is beyond the realm of the generators of $E_{10}$ considered in the eleven-dimensional analysis, but using the same, underformed sigma model. As a consequence, this work provides a dynamical unification of the massless and massive versions of type IIA supergravitiy inside $E_{10}$.
\end{abstract}
\maketitle                   





\section{Introduction}

An important class of supergravity theories is provided by deformed maximal supergravities, that are theories that cannot be obtained directly by standard toroidal Kaluza-Klein reduction of eleven-dimensional supergravity. They have as an important feature that they admit domain walls solutions, without which the duality symmetry of the underlying string theory cannot be verified. In particular, massive type IIA supergravity, unlike its massless sister, supports a D$8$-brane solution, that in IIA string theory can be reached from lower-dimensional branes by sequences of T-dualities~\cite{Polchinski:1995mt}. Therefore, any decription of M-theory should include deformed supergravities.

A possible approach to M-theory is via Kac-Moody symmetries, notably $E_{10}$
\cite{Julia:1980gr,Julia:1982gx,Julia:1997cy,Damour:2002cu} and $E_{11}$ \cite{Julia:1980gr,Julia:1982gx,Julia:1997cy,West:2001as,Englert1,Englert:2003py}. The $E_{10}$ proposal, on which we focus in this talk, has two main motivations. First the $E_{11-D}$ symmetry appearing in the reduction of eleven-dimensional supergravity to dimensions $D\geq 2$ naturally leads to the conjecture that the reduction to one dimension should be invariant under $E_{10}$ \cite{Julia:1980gr,Julia:1982gx}. Second, it is remarkable that the same intuition comes from cosmological billiards: close to a spacelike singularity (the BKL limit), eleven-dimensional supergravity becomes explicitly symmetric under the Weyl group of $E_{10}$~\cite{Damour:2002cu}.

Moreover, recent development observed the relevance of $E_{10}$ and $E_{11}$ in the framework of deformed supergravities, where the deformation parameters are identified with forms of high rank, specifically $(D-1)$-forms for
deformations in $D$ dimensions~\cite{Riccioni:2007au,Bergshoeff:2007qi,Riccioni:2007ni}.

The purpose of this talk is to explain how the deformation parameter of massive IIA supergravity enters the dynamics of the geodesic model of $E_{10}$, and our analysis includes the fermions. We show that the mass enters as the dual to a generator that is outside the realm of the generators considered usually. Importantly, all terms associated with the mass coincide perfectly. Here we focus on the general features and refer the reader to
~\cite{mainpaper}, upon which this talk is based, for the technical details.

\section{Massive IIA supergravity}
The first construction of massive type IIA supergravity is due to Romans \cite{Romans:1985tz} and its main step is to give a mass to the two-form potential of standard IIA supergravity~\cite{Giani:1984wc,Campbell:1984zc,Huq:1983im} through the replacement $F_{(2)}\rightarrow F_{(2)}+mA_{(2)}$, where $F_{(2)}=dA_{(1)}$. The one-form potential $A_{(1)}$ is then gauged away, which leads to terms depending on $m^{-1}$ in the supersymmetry variations and thus obscures the massless limit. This is remedied by a field redefinition presented in \cite{Bergshoeff:1996ui,Lavrinenko:1999xi}, that we employ in the analysis. A more democratic version of massive IIA supergravity is given in \cite{Lavrinenko:1999xi,Bergshoeff:2001}. It includes a nine-form dual to the mass $m\propto \ast_{10}dA_{(9)}$, and it is precisely that dual nine-form that we will be able to identify with a nine-form appearing in a certain decomposition of $E_{10}$. 

\begin{figure}
\begin{center}
\includegraphics[scale=0.8]{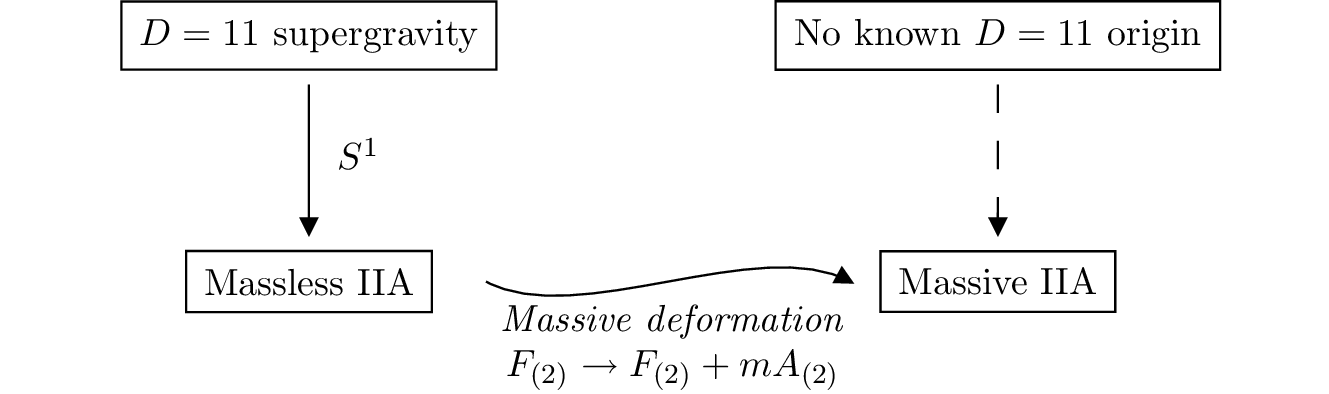}
\end{center}
\caption{\label{mIIA} \sl Massive IIA supergravity from $D=11$
    supergravity: Massive type IIA supergravity is obtained as a deformation of the standard type IIA supergravity, but unlike the latter, it does not possess any known eleven-dimensional origin. See Figure \ref{E10mIIA} for a pictorial description of how $D=11$ supergravity and massive IIA supergravity are unified inside $E_{10}$.}
\end{figure}

\begin{figure}[t!]
\begin{center}
\includegraphics[scale=0.8]{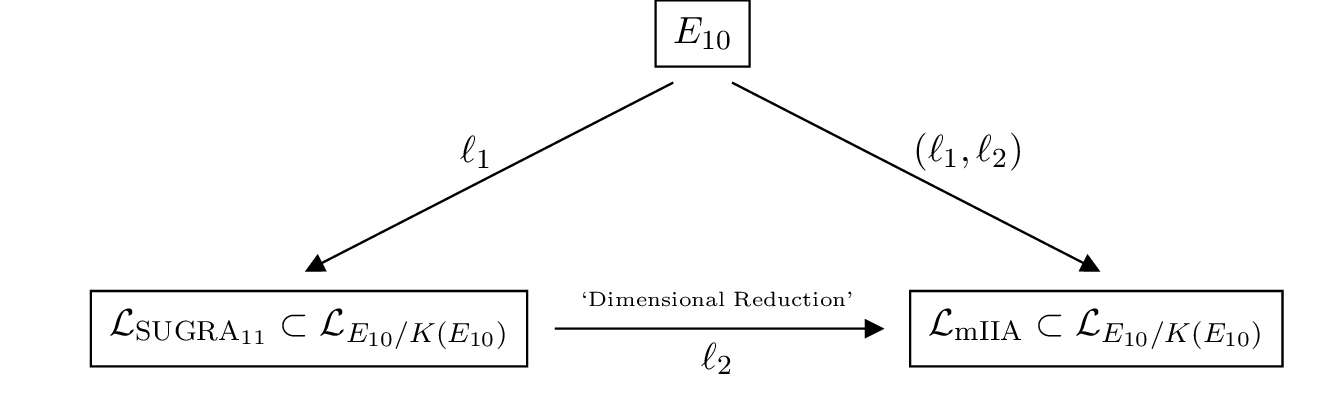}
\end{center}
\caption{\label{E10mIIA} \sl This picture describes the common $E_{10}$ origin
  of eleven-dimensional supergravity and massive type IIA supergravity. First,
  if one considers a level $\ell_1$ decomposition of $E_{10}$ with respect to
  $A_9$ (cf. Figure \ref{figure:E10}), one sees that the first levels
  ($\ell_1=0$ to $\ell_1=3$) of an $E_{10}/K(E_{10})$ sigma-model correspond
  to a truncated version of eleven-dimensional supergravity, with Lagrangian
  $\mc{L}_{\mathrm{SUGRA}_{11}}$ \cite{Damour:2002cu}. Taking this as a
  starting point, we can perform an additional level $\ell_2$ decomposition on
  the sigma model. On the lower $\ell_1$ levels ($\ell_1=0$ to $\ell_1=3$)
  this is equivalent to a dimensional reduction of eleven-dimensional
  supergravity, which gives massless IIA supergravity (cf. Figure
  \ref{mIIA}). However, if one includes one of the generators appearing at
  $\ell_1 = 4$, this leads to a theory that coincides with a truncated version
  of massive IIA supergravity, with Lagrangian
  $\mathcal{L}_{\mathrm{mIIA}}$. This procedure is equivalent to a multi-level
  $(\ell_1,\ell_2)$ decomposition of $E_{10}$ with respect to $A_8$.}  
\end{figure}

Moreover, massive IIA supergravity has in common with many other deformed maximal
supergravities that it does not possess any known higher-dimensional origin,
as illustrated in Figure \ref{mIIA}. A consequence of the present work is to
show that, although they are not related by dimensional reduction, eleven
dimensional supergravity and massive IIA supergravity have the same $E_{10}$
origin as displayed in Figure \ref{E10mIIA}, see
also~\cite{Damour:2002fz,Kleinschmidt:2003mf,West:2004st}.

In the form we consider in this work, the bosonic sector of massive IIA supergravity contains a
metric, a dilaton, a one-form, a two-form, a
three-form, and a real mass parameter
$m$. On the fermionic side, we have two gravitini, combined in a single $10\times32$ component
vector-spinor, and two dilatini, combined in a single $32$
component Dirac-spinor, which decompose into two
fields of opposite chirality under $SO(1,9)$. The full expression of the Lagrangian in our conventions is given in \cite{mainpaper}.

\section{$E_{10}$ and the geodesic sigma model for $E_{10}/K(E_{10})$}
\subsection{Generalities on $E_{10}$ and $K(E_{10})$.}
Here we summarize important features about the Kac-Moody algebras $\mf{e}_{10}$ and $\mf{k}(\mf{e}_{10})$, the groups of which we shall denote by $E_{10}$ and $K(E_{10})$. More details can be found in \cite{Damour:2002cu,bigreview,mainpaper}. 

\begin{figure}[t]
\centering
\includegraphics[scale=0.8]{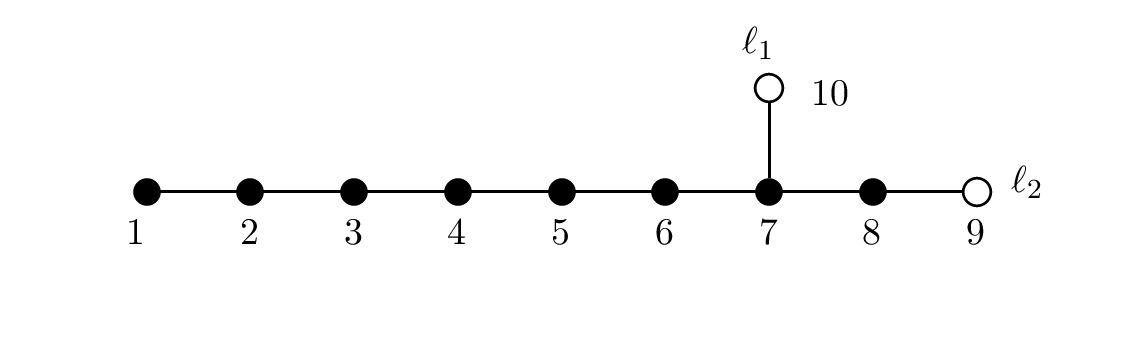}
\caption{\label{e10dynk}\sl The Dynkin diagram of $\mf{e}_{10}$ with the nodes associated with the level decomposition indicated in white.}
\label{figure:E10}
\end{figure}

The split real form of $\mf{e}_{10}$ is generated by ten triples $(e_i, f_i, h_i)$, $i=1,\dots ,10$, of Chevalley generators, each triple making up a distinguished $\mf{sl}(2, \mbb{R})$ subalgebra, 
These subalgebras are intertwined inside $\mf{e}_{10}$ according to the stucture of the Dynkin diagram in Figure \ref{figure:E10}. 

The maximal compact subalgebra $\mf{k}(\mf{e}_{10})\subset \mf{e}_{10}$ is defined as the subalgebra which is invariant under the Chevalley involution $\om$, which is defined through its action on each triple $(e_i, f_i, h_i)$:
\beq
\om(e_i)=-f_i, \qquad \om(f_i)=-e_i, \qquad \om(h_i)=-h_i.
\eeq
The subalgebra $\mf{k}(\mf{e}_{10})$ enters the so-called Iwasawa decomposition of $\mf e_{10}$,
\be
\label{iwasawa}
\mf{e}_{10} &=& \mf{k}(\mf{e}_{10})\oplus \mf{h} \oplus \mf{n}_+,
\ee
where $\mf{h}$ is the Cartan subalgebra, generated by the $h_i$ and $\mf{n}_+$ is the infinite-dimensional positive nilpotent subalgebra, generated by the positive step operators $e_i$.

\subsection{The $A_8$ level decomposition of $E_{10}$}

The correspondence between $\mf{e}_{10}$ and eleven-dimensional supergravity is made by introducing an $A_9\cong \mf{sl}(10, \mbb{R})$ level decomposition of $E_{10}$, where the level $\ell_1$ of a root $\alpha$ of $\mf e_{10}$ is its integer coordinate in the direction of the simple root $\alpha_{10}$ (associated to node $10$ in the Dynkin diagram in Figure \ref{figure:E10}) \cite{Damour:2002cu}. For each value of the level $\ell_1$, one has a finite number of representations of $A_9$. The correspondence was established up to level $\ell_1=3$ (with some minor exceptions \cite{Damour:2004zy}).

In the case we are interested in here, one needs to perform a further decomposition $\ell_2$ associated to the root $9$. Hence, we write any root $\alpha$ of $\mf e_{10}$ in terms of the ten simple roots as
\beq
\alpha=\ell_1\alpha_{10}+\ell_2\alpha_{9}+\sum_{i=1}^8m_i\alpha_i.
\eeq
The level $\ell:=(\ell_1,\ell_2)$ is now two-folded and corresponds to a decomposition under the $A_8\cong \mf{sl}(9, \mbb{R})$ subalgebra of $\mf{e}_{10}$, defined by nodes $1, \dots, 8$ in the Dynkin diagram in Figure \ref{figure:E10}. At level $(0,0)$, there is a copy of $\mf{gl}(9, \mbb{R})$, $K^a{}_b$, and a scalar generator, $T$, associated with the dilaton.
The generators of $\mf{e}_{10}$ at higher levels are $\mf{sl}(9, \mbb{R})$-tensors of higher and higher rank $E_{a_1\cdots a_k}\in \mf{e}_{10}$, where $k=2\ell_1+\ell_2$ and $a_i=1,\dots,9$. The full table up to $\ell=(4,1)$ can be found in \cite{mainpaper}.

In particular, at $\ell=(4,1)$, one has a nine-form generator $E^{a_1\cdots a_9}$ whose accompanying nine-form field will be identified with the dual to the mass of massive IIA supergravity \cite{Damour:2002fz,Schnakenburg:2002xx}. This is intriguing since in $D=11$ the matching between supergravity and $\mf{e}_{10}$ has only been successful up to $\ell_1=3$. Hence, the mass term in $D=10$ is outside this and provides a non-trivial check of $\mf{e}_{10}$ beyond its `$\mf{sl}(10, \mbb{R})$-covariantized $\mf{e}_8$' subset, i.e. the generators of $\mf{e}_8$ and their images under (the Weyl group of) $\mf{sl}(10, \mbb{R})$.

\subsection{Construction of the non-linear sigma model}

We here describe how to build the non-linear sigma model with rigid $E_{10}$ invariance and local $K(E_{10})$ invariance. Thanks to the Iwasawa decomposition (\ref{iwasawa}), one can choose a representative of the coset space $E_{10}/K(E_{10})$ in the so-called partial `Borel gauge' by taking only exponentials of $\mf{h}$ and $\mf{n}_+$:
\beq
\cV(t) = e^{h^a{}_b(t)K^b{}_a}   e^{\phi T} e^{A(t)\star E}  \hs \in \hs E_{10}/K(E_{10}),
\eeq
where $A(t)\star E$ is a sum over the positive level generators $E^{a_1\cdots a_k}$ of $E_{10}$ with coefficients $A_{a_1\cdots a_k}(t)$.
The coset representative $\mc{V}$ transforms under global $g\in E_{10}$-transformations from the right and local $k\in K(E_{10})$-transformations from the left $\mc{V} \longmapsto k\mc{V}g$. From $\cV(t)$, one can construct the Lie-algebra element in Maurer-Cartan form
\beq
v(t)=\p_t\cV\cV^{-1}= \cP(t) +\cQ(t),
\eeq
that decomposes, under the Chevalley involution, into an invariant part ($\cQ\in\mf{k}(\mf{e}_{10})$) and an anti-invariant part ($\cP\in\mf{e}_{10}\ominus\mf{k}(\mf{e}_{10})$).

In the next section, we will identify the fields of massive IIA supergravity with the components of $\cP(t)$ and $\cQ(t)$ in the $A_8$ level decomposition of $\mf e_{10}\ominus\mf{ke}_{10}$ or $\mf{ke}_{10}$ respectively, that we will note $P^{(\ell)}$ and $Q^{(\ell)}$ at level $\ell$. Because of the choice of the partial Borel gauge for $\cV$, $P^{(\ell)} =Q^{(\ell)},\,\, \, \forall\ell\neq(0,0)$.

\subsubsection{The bosonic part}
A manifestly $E_{10}\times K(E_{10})_{\text{local}}$-invariant Lagrangian is constructed as follows \cite{Damour:2002cu,Damour:2004zy}
\beq\label{e10boslag}
\mc{L}^{[B]}_{E_{10}/K(E_{10})}=\f{1}{4}n(t) ^{-1}\left<\mc{P}(t)\big|\mc{P}(t)\right>,
\eeq
where the bracket represents an invariant inner product over $E_{10}$ and the lapse function $n(t)$ ensures invariance under reparametrizations of the geodesic parameter $t$. The equations of motion for $\mc{P}$ (in the gauge $n=1$) read
\beq
\mc{D}\mc{P}:= \pa_t \mc{P}-[\mc{Q}, \mc{P}]=0,
\label{bosoniceom}
\eeq
where we defined the $K(E_{10})$-covariant derivative $\mc{D}$. 
\subsubsection{The fermionic part}

In order to build the fermionic part of the $E_{10}/K(E_{10})$ sigma model, one needs to introduce spinorial representations of $\mf{k}(\mf{e}_{10})$. In the case of eleven-dimensional supergravity, a good correspondence is obtained using two finite-dimensional (unfaithful) representations. The first one transforms as a $32$-dimensional Dirac-spinor representation $\epsilon$ of $\mf{so}(10)\subset \mf{k}(\mf{e}_{10})$ and corresponds to the supersymmetry parameter. The second one transforms as a $320$-dimensional vector-spinor representation  $\Psi_{\dot{a}}, \hs \dot{a}=(10, a)$, of $\mf{so}(10)\subset \mf{k}(\mf{e}_{10})$ and is identified with the gravitino \cite{deBuyl:2005zy,Damour:2005zs,deBuyl:2005mt,Damour:2006xu}. Upon reduction to the IIA theory (through the additional level decomposition with respect to $\ell_2$), while the supersymmetry parameter stays unchanged, the gravitino decomposes into a $32$-dimensional spinor $\Psi_{10}$ (to be associated with the ten-dimensional dilatino) and a 288-dimensional vector spinor $\Psi_a$ of $\mf{so}(9)$ (related to the gravitino) that will mix under $\mf{k}(\mf{e}_{10})$~\cite{Kleinschmidt:2006tm}. 

The fermionic degrees of freedom are included in the Lagrangian through the spinor representation $\Psi$ as follows \cite{Damour:2005zs,deBuyl:2005mt,Damour:2006xu}
\beq\label{e10fermlag}
\mc{L}^{[F]}_{E_{10}/K(E_{10})}=-\f{i}{2}\left<\Psi \big| \mc{D} \Psi\right>,
\eeq
where the bracket now denotes an invariant inner product on the representation space. The associated `Dirac equation' reads
\beq
\mc{D}\Psi:= \pa_t\Psi-\mc{Q}\cdot \Psi=0.
\label{fermioniceom}
\eeq

The bosonic equations of motion (\ref{bosoniceom}) were written for the gauge choice $n=1$. The lapse function $n$ has a superpartner $\Psi_t$, which is a Dirac spinor under $\mf{k}(\mf{e}_{10})$, as is the supersymmetry parameter, and the associated supersymmetry transformations are
\be \label{SusyTransfSigmaModel}
\delta_{\epsilon} n &=& i\epsilon^T\Psi_t,\nn \\
\delta_{\epsilon} \Psi_t &=& \mc{D}\epsilon.
\ee
The fermionic equations of motion are then valid in the `supersymmetric gauge' $\Psi_t=0$. 

\section{The correspondence}
In order to compare the equations (of motion and of supersymmetry) of supergravity to the equations of our sigma model, we need to rewrite the former. First, as is customary in the correspondence between $E_{10}$ and supergravity we split the
indices into temporal and spatial indices and adopt
a pseudo-Gaussian gauge for the ten-dimensional vielbein. In addition we demand that the spatial trace of the spin connection vanishes. We also choose temporal gauges for all supergravity gauge potentials. Moreover, we can only expect that a truncated version of the supergravity
equation corresponds to the coset model equations. This truncation was originally devised in the context of eleven-dimensional
supergravity, where it was strongly motivated by the billiard analysis of the
theory close to a spacelike singularity (the `BKL-limit')
\cite{Damour:2002cu,Damour:2004zy}. In this limit, spatial points decouple and
the dynamics becomes effectively time-dependent, ensuring that the truncation
is a valid one in this regime. In this paper, we analyse the same question in
the context of massive IIA supergravity, and an identical procedure requires
the truncation of a set of spatial gradients. These can be obtained from a
BKL-type analysis of massive IIA.

One can now proceed to the comparison between the two theories. In practice, we compare the equations as prescribed in Table \ref{comp}.
\begin{table}[t]
\centering
\begin{tabular}{|c|c|}
\hline
Supergravity&$E_{10}/K(E_{10})$ \\
\hline
Bianchi identities and bosonic equations of motion& $\pa_t \mc{P}-[\mc{Q}, \mc{P}]=0$\\[.2cm]
Fermionic equations of motion& $\pa_t\Psi-\mc{Q}\cdot \Psi=0$\\[.2cm]
Supersymmetry variation of $\psi_t$& $\delta_{\epsilon} \Psi_t = \mc{D}\epsilon$\\
\hline
\end{tabular}
\caption{\label{comp}\sl Each line of this table contains the equations to be compared to each other in order to make the correspondence between massive IIA supergravity and $E_{10}$ explicit.}
\end{table}

As a result, we obtain a dictionary between the bosonic and fermionic fields of massive IIA supergravity and the representations of $\mf{e}_{10}$ and $\mf{k}(\mf{e}_{10})$ that we defined in the previous section. The schematic correspondence is presented in Table \ref{corr}.
\begin{table}[t]
\centering
\begin{tabular}{|c|c|}
\hline
Supergravity&$E_{10}/K(E_{10})$ \\
\hline
Bosonic fields& $P^{(\ell)} (\ell\geq 0), Q^{(0)}$\\[.2cm]
Fermionic fields& $\Psi_t$, $\Psi_a$, $\Psi_{10}$\\[.2cm]
Supersymmetry parameter& $\epsilon$\\
\hline
\end{tabular}
\caption{\label{corr}\sl This table shows schematically which fields of the two theories are identified in the correspondence.}
\end{table}

This correspondence works perfectly up to level $(4,1)$ for all equations but one: the Einstein equation does not fit perfectly in this picture. More precisely, two terms do not match completely with the corresponding sigma model equation. These discrepancies can however be traced back to $D=11$ supergravity where
both mismatches were part of the $D=11$ Ricci tensor \cite{Damour:2004zy}. In this sense this is
not a new discrepancy but a known one. It is to be noted that all the terms
involved in the mismatch are related to contributions to the Lagrangian which
would give rise to walls corresponding to imaginary roots in the cosmological
billiards picture \cite{Damour:2002cu}.

Moreover, in particular, and most importantly, one notices that the mass enters all equations correctly when identified with the nine-form $P_{a_1\cdots a_9}$ of $E_{10}$ at level $(4,1)$ in the following way:
\beq
P_{a_1\cdots a_9}=\f12Ne^{5\phi/2}\eps_{a_1\cdots a_9}m,
\eeq
where $N$ is the lapse and $\phi$ the dilaton of massive IIA supergravity. 

Further aspects of gauge fixing and the consistency of the gauge algebra and supersymmetry with the correspondence can be found in \cite{mainpaper}.

\begin{acknowledgement}
E.J. would like to thank the organisers of the fourth RTN "Forces-Universe" Workshop 2008, for giving her the opportunity to give this talk. E.J. is a bursar of Fonds de la Recherche Scientifique--FNRS,
Belgium. A.K. is a Research Associate of the Fonds de la Recherche
Scientifique--FNRS, Belgium. Work supported in part by IISN-Belgium
(conventions 4.4511.06 and 4.4514.08), by the European Commission FP6 RTN
programme MRTN-CT-2004-005104 and by the Belgian Federal Science Policy Office
through the Interuniversity Attraction Pole P6/11. 
\end{acknowledgement}


\begin{thebibliography}{10}

\bibitem{Polchinski:1995mt}
  J.~Polchinski,
  Phys.\ Rev.\ Lett.\  {\bf 75} (1995) 4724
  [arXiv:hep-th/9510017].


\bibitem{Julia:1980gr} B.~Julia, Invited paper presented at Nuffield Gravity  Workshop,  Cambridge, Eng., June 22 - July 12, 1980.

\bibitem{Julia:1982gx} B.~Julia, in: Lectures in Applied Mathematics, AMS-SIAM {\bf 21} (1985) 335.

\bibitem{Julia:1997cy}  B.~L.~Julia, [arXiv:hep-th/9805083].

\bibitem{Damour:2002cu}
  T.~Damour, M.~Henneaux and H.~Nicolai,
  Phys.\ Rev.\ Lett.\  {\bf 89} (2002) 221601
  [arXiv:hep-th/0207267].

\bibitem{West:2001as}
  P.~C.~West,
Class.\ Quant.\ Grav.\  {\bf 18} (2001) 4443
  [arXiv:hep-th/0104081].

\bibitem{Englert1}
F.~Englert, L.~Houart, A.~Taormina and P.~West, JHEP {\bf 09} (2003) 020. [arXiv:hep-th/0304206].


\bibitem{Englert:2003py}
  F.~Englert and L.~Houart,
   JHEP {\bf 0401} (2004) 002
  [arXiv:hep-th/0311255].

\bibitem{Riccioni:2007au}
  F.~Riccioni and P.~C.~West,
  JHEP {\bf 0707} (2007) 063
  [arXiv:0705.0752 [hep-th]].

\bibitem{Bergshoeff:2007qi}
  E.~A.~Bergshoeff, I.~De Baetselier and T.~A.~Nutma,
  JHEP {\bf 0709} (2007) 047
  [arXiv:0705.1304 [hep-th]].

\bibitem{Riccioni:2007ni}
  F.~Riccioni and P.~C.~West,
  JHEP {\bf 0802} (2008) 039
  [arXiv:0712.1795 [hep-th]].

\bibitem{mainpaper}
M.~Henneaux, E.~Jamsin, A.~Kleinschmidt and D.~Persson,[arXiv:0811.4358]

\bibitem{Romans:1985tz}
  L.~J.~Romans,
  Phys.\ Lett.\  B {\bf 169} (1986) 374.
\bibitem{Giani:1984wc}
  F.~Giani and M.~Pernici,
  Phys.\ Rev.\  D {\bf 30} (1984) 325.
\bibitem{Campbell:1984zc}
  I.~C.~G.~Campbell and P.~C.~West,
  Nucl.\ Phys.\  B {\bf 243} (1984) 112.

\bibitem{Huq:1983im}
  M.~Huq and M.~A.~Namazie,
  Class.\ Quant.\ Grav.\  {\bf 2} (1985) 293
  [Erratum-ibid.\  {\bf 2} (1985) 597].
\bibitem{Bergshoeff:1996ui}
  E.~Bergshoeff, M.~de Roo, M.~B.~Green, G.~Papadopoulos and P.~K.~Townsend,
  Nucl.\ Phys.\  B {\bf 470} (1996) 113
  [arXiv:hep-th/9601150]

\bibitem{Lavrinenko:1999xi}
  I.~V.~Lavrinenko, H.~Lu, C.~N.~Pope and K.~S.~Stelle,
  Nucl.\ Phys.\  B {\bf 555} (1999) 201
  [arXiv:hep-th/9903057]
\bibitem{Bergshoeff:2001}
E.~Bergshoeff, R.~Kallosh, T.~Ortin, D.~Roest and A.~Van Proeyen,
Class.\ Quant.\ Grav. {\bf 18} (2001) 3359
[arXiv:hep-th/0103233]

  \bibitem{Damour:2002fz}
  T.~Damour, S.~de Buyl, M.~Henneaux and C.~Schomblond,
  JHEP {\bf 0208} (2002) 030
  [arXiv:hep-th/0206125].

\bibitem{Kleinschmidt:2003mf}
  A.~Kleinschmidt, I.~Schnakenburg and P.~C.~West,
  Class.\ Quant.\ Grav.\  {\bf 21} (2004) 2493
  [arXiv:hep-th/0309198].

\bibitem{West:2004st}
  P.~C.~West,
  Nucl.\ Phys.\  B {\bf 693} (2004) 76
  [arXiv:hep-th/0402140].
  
\bibitem{bigreview}
 M.~Henneaux, D.~Persson and P.~Spindel,
  Living Rev.\ Rel.\  {\bf 11} (2008) 1
  [arXiv:0710.1818 [hep-th]].
\bibitem{Damour:2004zy}
  T.~Damour and H.~Nicolai,
  in: G.~S.~Pogoyan, L.~E.~Vicent and
  K.~B.~Wolf (eds.), Group 
  Theoretical Methods in Physics (IoP Conference Series Number 185), IoP
  Publishing  (2005) 93
[arXiv:hep-th/0410245].


\bibitem{Schnakenburg:2002xx}
  I.~Schnakenburg and P.~C.~West,
  {\sl Massive IIA supergravity as a non-linear realisation},
  Phys.\ Lett.\  B {\bf 540} (2002) 137
  [arXiv:hep-th/0204207].
 
\bibitem{deBuyl:2005zy}
  S.~de Buyl, M.~Henneaux and L.~Paulot,
  Class.\ Quant.\ Grav.\  {\bf 22} (2005) 3595
  [arXiv:hep-th/0506009].
\bibitem{Damour:2005zs}
  T.~Damour, A.~Kleinschmidt and H.~Nicolai,
  Phys.\ Lett.\  B {\bf 634} (2006) 319
  [arXiv:hep-th/0512163].

\bibitem{deBuyl:2005mt}
  S.~de Buyl, M.~Henneaux and L.~Paulot,
  JHEP {\bf 0602} (2006) 056
  [arXiv:hep-th/0512292].
\bibitem{Damour:2006xu}
  T.~Damour, A.~Kleinschmidt and H.~Nicolai,
  JHEP {\bf 0608} (2006) 046
  [arXiv:hep-th/0606105].
\bibitem{Kleinschmidt:2006tm}
  A.~Kleinschmidt and H.~Nicolai,
  Phys.\ Lett.\  B {\bf 637} (2006) 107
  [arXiv:hep-th/0603205].




\end{thebibliography}
\end{document}